\documentstyle[12pt,epsfig]{article}
\newcommand{\be}{\begin{equation}}
\newcommand{\bea}{\begin{eqnarray}}
\newcommand{\eea}{\end{eqnarray}}
\newcommand{\ba}{\begin{array}}
\newcommand{\ea}{\end{array}}
\newcommand{\ee}{\end{equation}}

\newcommand{\tr}{{\rm Tr}}

\expandafter\ifx\csname mathbbm\endcsname\relax

\else

\fi
\begin{document}
\begin{titlepage}
\hfill
\vbox{
    \halign{#\hfil         \cr
           CERN-TH/2000-251 \cr
           hep-th/0008172  \cr
           } 
      }  
\vspace*{20mm}
\begin{center}
{\Large {\bf  Supergravity and ``New'' Six-Dimensional Gauge Theories}\\} 

\vspace*{15mm}
\vspace*{1mm}
{Mohsen Alishahiha and Yaron Oz}\\

\vspace*{1cm} 

{\it Theory Division, CERN \\
CH-1211, Geneva, 23, Switzerland}\\

\vspace*{1cm}
\end{center}

\begin{abstract}

In the first part of this letter, we analyse the supergravity dual descriptions of six-dimensional
field theories realized on the worldvolume of $(p,q)$ five-branes (OD5 
theory).
We show that in order for the low-energy gauge theory description
to be valid the $\theta$ parameter must be rational.
Irrational values of $\theta$ require a  strongly coupled string 
description of the system 
at low-energy.
We discuss the phase structure and deduce some properties
of these theories.
In the second part we construct and study  the supergravity description of NS5-branes with
two
electric RR field,
which provides a dual description of 
six-dimensional theories with several light open D-brane excitations.

\phantom{\cite{AOR,AOJ,nil,MR,RS,WU,LRO,PAP,ILPT,CP,OHZ,OHTA}}

\end{abstract}
\vskip 5cm

August 2000
\end{titlepage}

\newpage

\section{ Introduction}

This letter consists of two parts.
In the first part we  
will consider the $(1,1)$ supersymmetric six-dimensional gauge
theories on the worldvolume of $(p,q)$ five branes in Type IIB string theory
introduced in \cite{Witten}.
These theories have been reconsidered in the context of non-commutative
deformation of the NS5-brane theory \cite{GMSS,har} and are named
OD5 theories.
At low energy the theories reduce to super Yang-Mills (SYM) theories
with gauge group $SU(s)$ where $s$ is the greatest common divisor of $p$
and $q$.  

While the low energy interactions dominated by the gauge kinetic
 term cannot distinguish
the different  $(p,q)$ theories with the same low-energy gauge group,
it was argued in \cite{Witten} that
the $\theta$ term related to the 
$\pi_5$ homotopy group of $SU(s)$ is an observable
that distinguishes the low-energy gauge theories.
One issue which is not settled is what values of $\theta$
are possible in the low-energy theories.
The construction in \cite{Witten} considered rational values
of the $\theta$ parameter. A generalization to irrational $\theta$ was 
suggested in \cite{Kol}.     

In order to study this question we will consider the 
dual string (supergravity) description.
We will show that in order for the low-energy gauge theory description
to be valid $\theta$ must be rational.
Irrational values of $\theta$ require a strongly
coupled Type IIB string description of the system
at low-energy.
We will discuss the phase structure and deduce some properties
of these theories.

In the second part of the letter we will
construct and analyse the supergravity description of NS5-branes with
two electric RR field. This corresponds to a generalization
of the ODp theories to six-dimensional theories
with several types of light open D-branes. 
We will find two interesting cases.
The first case corresponds to $N$ Type IIB NS5-branes in the presence of 
electric RR 2-form  and 6-form potentials.
The background contains (D(-1),D1,D3,D5)-branes charges, and the six-dimensional worldvolume
theory contains the corresponding  D-branes.
In addition the background contains F-string and NS5-branes charges.
This is analogous
to Type IIB string theory, namely the 
six-dimensional theory contains all the possible branes up to dimension six.
All together, all the background fields have their  
$SL(2,Z)$ counterparts. 
This points to an $SL(2,Z)$ symmetry of the six-dimensional theory, much like Type IIB string theory.

The second case corresponds to $N$ Type IIA NS5-branes in the presence of 
electric RR 3-form  and 5-form potentials.
The background contains (D0,D2,D4)-branes charges, and the six-dimensional worldvolume
theory contains the corresponding  D-branes.
In addition the background contains F-string and NS5-branes charges.
This is analogous
to Type IIA string theory, namely the 
six-dimensional theory contains all the possible branes up to dimension six.

On the supergravity side  these will relate and generalize other 
constructions in \cite{AOR}--\cite{OHTA}.

The letter is organized as follows. In section 2 we will analyse the 
 $(p,q)$ theories from field theory and supergravity points of view and analyse
the possible values of the low-energy
$\theta$ parameter.
In section 3 we will construct 
the supergravity description of NS5-branes with
two electric RR field and study the corresponding 
six-dimensional theories with several types of open D-branes.

\section{``New'' Six-Dimensional Gauge Theories}

\subsection{Field Theory}

We will consider the $(1,1)$ supersymmetric six-dimensional gauge
theories on the worldvolume of $(p,q)$ five branes in Type IIB string theory.
At low energy the theories reduce to super Yang-Mills (SYM) theories
with gauge group $SU(s)$ where $s$ is the greatest common divisor of $p$
and $q$.  
Consider the low-energy effective action 
\bea
S_{eff} = \int {\rm d}^6x\left( \frac{1}{g_{YM}^2}\tr F^2 + \theta_{YM} \tr F \wedge F \wedge F +
...\right) \ ,
\label{action}
\eea
where the dots correspond to higher dimension operators as well as the supersymmetric
completion.
Clearly, the low energy interactions dominated by the $F^2$ term cannot distinguish
the different  $(p,q)$ theories with the same low-energy gauge group.
However, it was argued in \cite{Witten} that
the theta term in (\ref{action}), which is  a consequence
of $\pi_5(SU(s)) = {\bf Z}, s > 2$, is an observable
that distinguishes the low-energy gauge theories.
The construction in \cite{Witten} considered rational values
of the $\theta_{YM}$ parameters. A generalization to irrational values of
$\theta_{YM}$ was suggested
in \cite{Kol}.     

In the following we will construct a dual string (supergravity) description
of these theories and analyse the possible values of $\theta_{YM}$.
We will then use the dual description to study some properties
of the theories.

\subsection{The Dual String (Supergravity) Description}

Consider a system of Type IIB NS5-branes in the presence of an electric RR 
6-form potential.
This branes system can be considered as a coinciding (D5,NS5) branes system. 
Recently, this theory has been reconsidered in the context of non-commutative
deformation of the NS5-brane theory \cite{GMSS,har}
and was called OD5 theory. 
We denote the worldvolume coordinates by $x_0,...,x_5$.

The supergravity 
background of the $q$ NS5-branes in the presence of an electric 
RR 6-form reads
\bea
ds^2&=&h^{-1/2}\left[-dx_0^2+\sum_{i=1}^{5}dx_i^2 +f(dr^2+r^2d\Omega^2_3)
\right] \ , \cr&&\cr
f&=&1+\frac{ q l_s^2}{2\cos\gamma\;r^2}\ ,\;\;\;\;\;\;\;\;\;\;\;
h^{-1}=\sin^2\gamma f^{-1}+\cos^2\gamma \ ,\cr
&&\cr
\chi&=&{\tan\gamma\over g_s}f^{-1}h+\chi_0\ ,\;\;\;\;\;\; \;l_s^{-2}
dA
=q\left(5{\tan\gamma\over g_s}f^{-1}h-\chi_0\right)\;\epsilon_3\ , 
\cr
&&\cr
e^{2\phi}&=&g_s^2fh^{-2}\ , \;\;\;\;\;\;\;\;\;\;\;\;\;\;\;\;\;\;\;\; 
l_s^{-2}dB =q\; \epsilon_3\ .
\label{OD5}
\eea
Here $\chi$ is the RR scalar, $A$ is the 2-form dual to 
RR 6-form, and $\epsilon_3$
is the volume 3-form of the 3-sphere.
$B_{\alpha\beta}$ is the NS 2-form 
potential.
Note that in comparison to the solution presented
in \cite{AOR} we added a constant $\chi_0$. This constant will be relevant 
later for the
discussion 
of the $\theta$ parameter and its possible values. 
The Type IIB complexified coupling $\tau$ combines the dilaton
and the RR scalar $\tau=ie^{-\phi}+\chi$.

The requirement of having $p$ D5-branes charge implies the condition
\be
5{\tan\gamma\over g_s}-\chi_0={p\over q} \ .
\label{cond}
\ee
Note that the background depends on four independent parameters, which we can choose to
be $(p,q,g_s,\chi_0)$.

We will now consider the limit in which this system of $(p,q)$ five branes
decouples from the bulk \cite{GMSS, har, AOR}.
In this limit we send  
$l_s\rightarrow 0$ and
keep 

\def\af{ \alpha'_{\rm eff}}
\def\lef{ l_{\rm eff} }
\be
\lef^2 \equiv {l_s^2\over \cos\gamma}\ ,\;\;\;\;\;u\equiv {r\over l_s^2}\ ,\;\;\;\;\;
{\tilde g}\ \lef^{2}\equiv g_sl_s^{2} \ ,
\label{delimit}
\ee  
fixed. This requires $g_s \rightarrow \infty$.

In this limit the supergravity solution (\ref{OD5}) reads
\bea
l_s^{-2}ds^2&=&{\hat h}^{-1/2}\left[-d{\tilde x}_0^2+\sum_{i=1}^{5}
d{\tilde x}_i^2+{ q \over u^2}(du^2+u^2d\Omega^2_3)\right]\ ,\cr
&&\cr
\chi&=&{1\over {\tilde g}} a_{\rm eff}^2u^2 {\hat h}+\chi_0
\ , \;\;\;\;\;\;\;\;\;\;\;\;\;\;\;\;\;\;
l_s^{-2}dB = q\;\epsilon_3 \ ,\cr
&&\cr
l_s^{-2}dA &=&q\left( {5a_{\rm eff}^2u^2\over 
{\tilde g}} {\hat h}-\chi_0\right)
\epsilon_3 \ ,\;\;\;\;\;\;\;
e^{2\phi}={\tilde g}^2 \frac{{\hat h}^{-2}}{a_{\rm eff}^2u^2}\ , 
\label{deod5}
\eea
where ${\hat h}^{-1}=1+a_{\rm eff}^2u^2$. we have also rescaled 
the worldvolume coordinates
${\tilde x}_{0,\cdots,5}={1\over \lef }x_{0,\cdots,5}$, and
\bea
a_{\rm eff}^2 \equiv \frac{2l_{\rm eff}^2}{q} \ .
\eea 
In this limit the condition (\ref{cond}) reads
\be
\frac{5}{{\tilde g}}={p\over q}+\chi_0 \ .
\label{eq}
\ee

At the energy range  $a_{\rm eff}u\ll 1$ the RR scalar is  $\chi = \chi_0$, 
and the theta angle is $\theta = 2 \pi \chi_0$.
Note, however that in the extreme IR (and extreme UV) the effective string 
coupling is large and we have to perform a duality transformation in order 
to obtain a weakly coupled description. This is the reason why we cannot 
identify the low energy Yang-Mills theta angle (\ref{action}) with  
$2 \pi \chi_0$.
Here, the question of what values of theta angles are possible in the low 
energy SYM theory is 
formulated as what values of $\chi$ are possible in a weakly coupled
gauge field theory description at small $u$. 
To answer this question
we will have to discuss
the phase structure of the theories and in particular the
regions of validity of the different descriptions.

The curvature of the metric (\ref{deod5}) reads
\be
l_s^2 {\cal R} = \frac{6}{q}\frac{1 - 2a_{\rm eff}^2 u^2}
{({1+a_{\rm eff}^2u^2})^{3/2}} \ ,
\label{CUR}
\ee 
which is small for a large number of NS5-branes $q$.

Consider the energy regime 
$a_{\rm eff}u\ll 1$. The background (\ref{deod5}) simplifies 
\bea
l_s^{-2}ds^2&=&\left[-d{\tilde x}_0^2+\sum_{i=1}^{5}
d{\tilde x}_i^2+{ q\over u^2}(du^2+u^2d\Omega^2_3)\right]\ ,\cr
&&\cr
\chi&=&\chi_0\ ,
\;\;\;\;\;\;\;\;\;\;\;\;\;\;\;\;\;\;
l_s^{-2}dA =-q\chi_0\;\epsilon_3 \ ,
\cr
&&\cr
e^{\phi}&=&\frac{{\tilde g}}{a_{\rm eff}u}\ , 
\;\;\;\;\;\;\;\;\;\;\;\;\;\;l_s^{-2}dB =q\;\epsilon_3 \ .
\label{sg}
\eea

In order for the supergravity description (\ref{sg}) to be valid
we require  the effective string coupling $e^{\phi} \ll 1$.
This condition is satisfied when $a_{\rm eff}u \gg {\tilde g}$, which in this energy
range implies that
${\tilde g}\ll 1$. This condition is not satisfied, in particular, in the extreme
IR regime where we expect a perturbative SYM field theory description of the system.
Since
the effective string coupling is large, 
we have to find an appropriate weakly coupled dual description.
In our case, an S-duality transformation
$\tau \rightarrow -\frac{1}{\tau}$ does not make the effective string coupling small
because of the presence of the RR scalar $\chi= \chi_0$.

We distinguish two cases. In the first case $\chi_0$ is rational and in the second
 $\chi_0$ is irrational.
Let us start with the first case.
We can make the  effective string coupling small
by the  use of a more general $SL(2,Z)$ transformation  
\be
\tau\rightarrow \frac{a\tau+b}{c\tau+d}\ ,\;\;\;\;
B\rightarrow dB-cA\ , \;\;\;\;\;A\rightarrow -bB+aA
\label{SL}
\ee
where  $ad-bc=1$ and $c\chi_0+d=0$.
The last requirement can be satisfied since $\chi_0$ is rational.
Note that with this transformation the NS 2-form potential is mapped to zero.
Thus, the five branes charges are mapped $(p,q) \rightarrow (s,0)$ where $s$
is the greatest common divisor of $p$ and $q$ \footnote{More precisely,
$s=\frac{q}{c}$ which is the greatest common divisor of $p$ and $q$ upon comparing
the $\theta$ angles in IR and UV.}, and the RR scalar is mapped
to the rational number $\chi = \frac{a}{c}$.

Under this $SL(2,Z)$ transformation we have 
\bea
g'_s={c^2\over g_s},\;\;\;\;\;{l'}_s^2={g_sl_s^2\over c} \ ,
\eea
and therefore

\bea
{\tilde g}'={c^2\over {\tilde g}},\;\;\;\;\;{\lef'}^2={{\tilde g}\lef^2
\over c} \ .
\eea

The dilaton maps to
\be
e^{2\phi'}=
\frac{{\tilde g}^{\prime 3}{\lef'}^2 {\tilde{u}}^2}{s} \ .
\ee
The metric and the dilaton take the familiar form of the D5-branes background
\bea
ds^2 &=& l_s^{\prime 2}\left(\frac{u}{(g_{YM}^2 s)^{1/2}}
(-d x_0^2+\sum_{i=1}^{5}
d x_i^2)+ \frac{(g_{YM}^2 s)^{1/2}}{u}
(du^2+u^2d\Omega^2_3)\right) \ , \nonumber\\
e^{\phi'} &=& \frac{(g_{YM}^2 s u^2)^{1/2}}{s} \ ,
\eea
where 
\bea
g_{YM}^2 \equiv  {\tilde g}^{\prime 3}{\lef'}^2 \ ,
\eea
and we have rescaled $\tilde{u} \rightarrow {u\over c}$.
The coordinates $x_i$ are the same as the ones in (\ref{OD5}).

Defining the dimensionless effective Yang-Mills coupling
by $g_{{\rm eff}}^2 \equiv g_{YM}^2 s u^2$, we recast the curvature and dilaton   
in the form \cite{malda}
\bea
l_s^{\prime 2}{\cal R} &\sim&
{1\over g_{\rm eff}} \ , \nonumber\\
e^{\phi'} &\sim& \frac{g_{\rm eff}}{s} \ .
\eea

When $g_{{\rm eff}} \ll 1$ we have a good description of the system as a perturbative 
$SU(s)$ SYM theory with a rational $\theta_{YM}$ term.
This is valid at low energies $u \ll \frac{1}{(g_{YM}^2 s)^{1/2}}$.

Consider now the cases when 
$\chi_0$ is irrational.  Since the equation
$c\chi_0+d=0$ is not satisfied
for integers $c$ and $d$, we cannot use $SL(2,Z)$
transformations in order to find a background with a 
small dilaton \footnote{Since at the supergravity level 
the symmetry group is  $SL(2,R)$, one could try to satisfy the condition
$c\chi_0+d=0$ by taking non-integer $c,d$. However, with such a transformation we will end up with
non-integer D5-branes charge.}.   
Thus, we must remain at low-energy
with a description where the effective string coupling is large and
the curvature small (for large $q$).
In particular, in this framework we do not have at low-energy a 
perturbative 
$SU(s)$ SYM theory with irrational $\theta_{YM}$ term.

Let us turn to the UV regime $a_{\rm eff}u\gg 1$. Here
the effects of nonzero RR fields are important.
Again, in order for the supergravity description to be valid
we require  $e^{\phi} \ll 1$.
This condition is satisfied 
when $a_{\rm eff}u \ll {\tilde g}^{-1}$, which in this energy
range implies that
${\tilde g}\ll 1$. As before, when
this condition is not satisfied
we would like to make the  effective string coupling small
by the  use of an $SL(2,Z)$ transformation.

In this limit the RR scalar reads
\be
\chi={1\over 5}\;{p\over q}+{6\over 5}\chi_0 \ .
\ee
If $\chi_0$ is rational, we can make the effective string coupling
small by applying the same $SL(2,Z)$ transformation used in the IR regime.
With this transformation, the condition $c\chi+d=0$ leads to $cp-dq=0$, or $p=sd$.
The NS 2-form potential is mapped to zero while
the RR fields and dilaton map to
\be
\chi'={a\over c},\;\;\;\;\;\;{l'}_s^{-2}dA'=s\epsilon_3,\;\;
\;\;\;\;
e^{2\phi'}=\frac{s{\tilde g}'}{(\lef'u)^2} \ .
\label{g2}
\ee
The metric reads
\be
ds^2=\left[-dx_0^2+\sum_{i=1}^{5}dx_i^2+\frac{R^2}{u^2}
(du^2+u^2d\Omega_3^2)\right] \ ,
\label{g1}
\ee
where $R^2 = s{\tilde g}'{\lef'}^2$.

Consider now a graviton scattering in the background (\ref{g2}),(\ref{g1}).
Let the graviton be polarized along the worldvolume of the five branes $\Psi=e^{i\omega t}\psi(u)$
\footnote{This 
analysis is similar to that done for NS5-branes
in the presence of RR 2-form or 3-form \cite{{AIO},{ALI}}.}. 
$\psi(u)$ satisfies the
differential
 equation 
\be
{\partial^2\psi \over \partial u^2}+{3\over u}\;{\partial\psi \over 
\partial u}+{\omega^2 R^2\over u^2}\psi=0 \ .
\label{equat}
\ee
The solutions to equation (\ref{equat}) take the form $u^{\alpha}$, where
\be
\alpha=-1\pm\sqrt{1-\omega^2 R^2} \ .
\ee
When $\omega R > 1$, $\alpha$ becomes imaginary and
we get a 
wave-like solution. This means that there is a nonzero absorption cross section
(in the decoupling limit) for a graviton scattered on the five branes with the energy 
$E > \frac{1}{R}$. 

We can recast
equation (\ref{equat}) as
\be
-{\partial^2\over \partial u^2}g(u)+v(u)g(u)=0 \ ,
\ee
where $g(u)=u^{3/2}\psi(u)$ and
\be
V(u)=({3\over 4}-
\omega^2 R^2){1\over u^2} \ .
\label{pot}
\ee
The shape of the potential (\ref{pot})
changes at $\omega R = {\sqrt{3}\over 2}$.
 
As in \cite{MS}, here the theory has a mass gap 
\be
M_{\rm gap} \sim {1\over {(s{\tilde g}'{\lef'}^2)^{1/2}}} \ .  
\ee
In comparison with the ordinary NS5-branes \cite{MS} 
where we have $M_{\rm gap} \sim 1/\sqrt{sl_s^2}$ which 
corresponds to a string excitation, 
here
the mass gap is of the order of
a the D1-branes tension. Note that
this is also the case for D5-branes in the presence of a rank four B field
\cite{AOJ}.

\section{NS5-branes with two electric RR-fields}

The theories of Type II NS5-branes in the presence of an electric RR $(p+1)$-form 
(ODp) are six dimensional theories with open Dp-branes. The supergravity 
description of these has been recently considered in \cite{AOR}. 
The background of NS5-branes  in the presence of an electric RR  
$(p+1)$-form has also a magnetic RR $(5-p)$-form.
In this section we will generalize this and  study Type II NS5-branes in the 
presence of two electric RR fields, which corresponds to six-dimensional theories
with several types of open D-branes. 
This can be done, for instance, by
starting with the Type IIB background
of D5-branes in the presence of a rank four B field and using various $T,S$ dualities \footnote{Starting with  the Type IIB background
of D5-branes in the presence of a rank six B field does not lead to new
solutions.}.
There are four classes of solutions that we get:\\

(A) Two electric RR $(p+1)$-forms potentials, $p=1,2,3,4$.\\

(B) Electric RR $(p+1)$ and $(p+3)$-forms, $p=0,1,2,3$, with a 
magnetic B field.\\ 

(C) Electric RR $(p+1)$ and $(5-p)$-forms, $p=0,1,2$.\\ 

(D) Electric RR $(p+1)$ and $(7-p)$-forms, $p=1,2,3$, with an 
electric B field.\\ 

Note that for all these backgrounds, for each $q$-form potential there is also
a $(6-q)$-form potential. 
After taking the decoupling limit, the backgrounds $(A),(B),(C)$ are 
dual to theories which are already known.
Cases (A) and (B) correspond to the ODp theories 
(with extra fields). 
Case
(C) corresponds to the theory of NS5-branes in the presence of a light-like RR field
\cite{AOR}. 
Case (D) is new and 
it is the aim of this
section to analyse  it.

The supergravity background of $N$ NS5-branes in the presence of
electric RR $(p+1)$ and $(7-p)$-forms, $p=1,2,3$, reads
\footnote
{The supergravity solution  when $p=1$ has already been constructed in \cite{LRO}.}
\bea
ds^2=(h_1h_2)^{-1/2}\left(-dx_0^2+dx_1^2+h_1\sum_{i=2}^{6-p}dx_i^2+
h_2\sum_{j=7-p}^{5}dx_j^2 
+f\left(dr^2+r^2d\Omega_3^2\right)\right) 
\label{metric}
\eea
where the functions $f$ and $h_i$ are given by \footnote{Note that we have changed
the normalization of fields in comparison to section 2.}
\bea
f=1+\frac{Nl_s^2}{\cos\theta_1\cos\theta_2 r^2}\ , \;\;\;\;\;\;\;\;\; 
h^{-1}_i=\sin^2\theta_if^{-1}+\cos^2\theta_i \ .
\eea 
In addition we have RR fields $A$ and an electric $B$ field 
\bea
A^{(5-p)}_{2\cdots (6-p)}&=&{\tan\theta_1\over g_s}f^{-1}h_1 \ , \;
\;\;\;\;\;\;\;\;\;\;\;\;\;\;\;
A^{(p-1)}_{(7-p)\cdots 5}={\tan\theta_2\over g_s}f^{-1}h_2 \ , \cr
&&\cr
-A^{(p+1)}_{01(7-p)\cdots 5}&=&\frac{\sin\theta_1\cos\theta_2}{g_s}f^{-1}h_2 \ ,
\;\;\;\;\;\;\;
A^{(7-p)}_{01\cdots (6-p)}=-\frac{\sin\theta_2\cos\theta_1}{g_s}f^{-1}h_1 \ , \cr
&&\cr
B_{01}&=&\sin\theta_1\sin\theta_2 f^{-1} \  , \;\;\;\;\;\;\;\;\;\;\;\;\;\;\;\;\;\;\;
e^{2\phi}=g_s^2fh_1^{-(p-1)/2}h_2^{(p-5)/2} \ .
\label{fields}
\eea

In the decoupling limit 
we take $l_s\rightarrow 0$ and keep 
\be
u \equiv {r\over l_s},\;\;\;\;\; b_i \equiv {l_s\over \cos\theta_i},\;\;\;\;\; g_sl_s \equiv
{\tilde g}(b_1b_2)^{1/2} \ ,
\ee
fixed.
In this limit the metric (\ref{metric}) reads 
\bea
l_s^{-2}ds^2=({\hat h}_1{\hat h}_2)^{-1/2}\left(-d{\tilde x}_0^2+
d{\tilde x}_1^2+{\hat h}_1\sum_{i=2}^{6-p}d{\tilde x}_i^2
+{\hat h}_2\sum_{j=7-p}^{5}d{\tilde x}_j^2
+{N\over u^2}\left(du^2+u^2d\Omega_3^2\right)\right) \ ,
\label{metric2} 
\eea
and the fields (\ref{fields}) are
\bea
A^{(5-p)}_{2\cdots (6-p)}&=&{l_s^{(5-p)}\over {\tilde g}}({b_1\over b_2})^{
(p-4)/2}a_1^2u^2{\hat h}_1 \ ,\;\;\;\;\;\;\;\;\;
A^{(p-1)}_{(7-p)\cdots 5}={l_s^{p-1}\over {\tilde g}}({b_1\over b_2})^{(p-2)/2}
a_2^2u^2{\hat h}_2 \ ,\cr
&&\cr
-A^{(p+1)}_{01(7-p)\cdots 5}&=&\frac{l_s^{(p+1)}}{{\tilde g}}({b_1\over b_2})^{
p/2}a_2^2u^2{\hat h}_2 \ ,
\;\;\;\;\;\;\;\;\;
A^{(7-p)}_{01\cdots (6-p)}=-\frac{l_s^{(7-p)}}{{\tilde g}}({b_1\over b_2})^{
(p-6)/2}a_1^2u^2{\hat h}_1 \ ,\cr
&&\cr
e^{2\phi}&=&{\tilde g}^2\frac{(b_1/b_2)^{3-p}}{a_1a_2u^2}\;
{\hat h}_1^{-(p-1)/2}{\hat h}_2^{(p-5)/2} \ ,
\;\;\;\;\;B_{01}=l_s^2a_1a_2u^2 \ .
\label{fields2}
\eea
Here $a_1=\frac{b_1}{Nb_2}, a_2=\frac{b_2}{Nb_1}, {\hat h}^{-1}_i=1+a_i^2u^2$,
and we have rescaled the coordinates 
\be
{\tilde x}_{0,1}={1\over (b_1b_2)^{1/2}}x_{0,1},\;\;\;\;
{\tilde x}_{2,\cdots,(6-p)}={b_1^{1/2}\over b_2^{1/2}l_s}x_{2,\cdots,(6-p)},
\;\;\;\;
{\tilde x}_{(7-p),\cdots,5}={b^{1/2}_2\over b_1^{1/2}l_s}x_{(7-p),\cdots,5} \ .
\ee

In the following we will consider the different cases $p=1,2,3$.\\

\underline{{\bf $p=1$}}\\

This case corresponds to $N$ Type IIB NS5-branes in the presence of 
electric RR 2-form  and 6-form potential.
The background contains (D(-1),D1,D3,D5)-branes charges, and the six-dimensional worldvolume
theory contains the corresponding  D-branes.
In addition the background contains F-string and NS5-branes charges.
This is analogous
to Type IIB string theory, namely the 
six-dimensional theory contains all the possible branes up to dimension six.
All together, all the background fields have their  
$SL(2,Z)$ counterparts. 
This points to an $SL(2,Z)$ symmetry of the six-dimensional theory, much like Type IIB string theory.

Such an $\left((F,D1),(NS5,D5)\right)$ 
 system has been also considered in \cite{LRO}. If we denote the
F-string, D1-branes, NS5-branes and D5-branes charges by the 
integers $N', M', N, M$ 
respectively, there is a relation \cite{LRO}
\be
(N,M)=s(c,d),\;\;\;\;\;\;\;(N',M')=s'(-d,c) \ ,
\ee
where the $(c,d)$ and $(s,s')$ pairs are relatively prime.

Consider the supergravity background (\ref{metric2}), (\ref{fields2}). 
The dimensionless deformation parameters are
of the form 
$a_{i} u, i = 1,2$.
Assume for simplicity that $a_1=a_2=a_{eff}$. The supergravity
background reads  
\bea
l_s^{-2}ds^2&=&(1+a^2_{\rm eff}u^2)\left
[-d{\tilde x}_0^2+d{\tilde x}_1^2+\frac{\sum_{i=1}^{5}d{\tilde x}_i^2}
{1+a^2_{\rm eff}u^2}+
{N\over u^2}\left(du^2+u^2d\Omega_3^2\right)\right]\cr
&&\cr
A_{2345}&=&{l_s^{4}\over {\tilde g}}
A,\;\;\;\;\;\;\;\;\;\;\;\;\;\;\;
\chi={1\over {\tilde g}}A,\cr
&&\cr
A_{01}&=&-\frac{l_s^{2}}{{\tilde g}}
A,
\;\;\;\;\;\;\;\;\;\;\;\;
A_{012345}=-\frac{l_s^{6}}{{\tilde g}}
A,\cr
&&\cr
B_{01}&=& l_s^2a_{\rm eff}^2u^2,\;\;\;\;\;\;\;\;\;\;
e^{2\phi}={\tilde g}^2a_{\rm eff}^2u^2A^{-2}
\label{back}
\eea
where 
\bea
A=a_{\rm eff}^2u^2/(1+a_{\rm eff}^2u^2) \ .
\label{A}
\eea 

Similarly to the OD5 theory considered in the previous section,
the background contains a RR scalar $\chi$. As done there, 
we can add to $\chi$ a constant $\chi_0$, which will also modify the 
RR 2-form and 6-form.
Note also that the dilaton in (\ref{back}) 
is exactly 
the same as that of the OD5 theory, and is large 
in the IR and UV regimes.
Again due to the 
RR scalar, we have to make use of the $SL(2,Z)$ symmetry in order to pass to 
a weakly coupled dual description. 

In the IR, the theory is similar to the  OD5.
We can neglect the RR potentials. However, the RR scalar
$\chi=\chi_0$ implies a specific form of the 
$SL(2,Z)$ symmetry and the analysis is identical to that performed in the
previous section leading to the same conclusions regarding  the low-energy
$\theta$ parameter.

In the UV regime the theory differs from the OD5 theory.
Now we have other RR fields as well as an electric
B field. 
The
dilaton is large. Since the
RR scalar is rational we can use the $SL(2,Z)$ symmetry to map the 
solution to a 
weakly coupled one, with
RR 6-form charge as well
as an electric B field.
This structure is very reminiscent of the
NCOS theory, now in six-dimensions.
Note that the most general background depends on five independent parameters, which
we can choose to be $(M,N,g_s,\chi_0,n)$ where $n$ is the D3-branes charge.

Finally, we note that the conjecture
of the self-duality of OD1 theory \cite{GMSS} is a special case of 
the $SL(2,Z)$ symmetry of this theory in the  energy range 
$a_1 u \ll 1$.\\


\underline{{\bf $p=2$}}\\

This case corresponds to $N$ Type IIA NS5-branes in the presence of 
electric RR 3-form  and 5-form potentials.
The background contains (D0,D2,D4)-branes charges, and the six-dimensional worldvolume
theory contains the corresponding  D-branes.
In addition the background contains F-string and NS5-branes charges.
This is analogous
to Type IIA string theory, namely the 
six-dimensional theory contains all the possible branes up to dimension six.

Consider the supergravity background (\ref{metric2}), (\ref{fields2}). 
The dimensionless deformation parameters are
of the form 
$a_{i}u, i,j=1,2$.
Assume for simplicity that $a_1=a_2=a_{eff}$. The supergravity
background reads

\bea
l_s^{-2}ds^2&=&(1+a^2_{\rm eff}u^2)\left
[-d{\tilde x}_0^2+d{\tilde x}_1^2+\frac{\sum_{i=1}^{5}d{\tilde x}_{i}^2}
{1+a^2_{\rm eff}u^2}+
{N\over u^2}\left(du^2+u^2d\Omega_3^2\right)\right] \ ,\cr
&&\cr
A_{234}&=&{l_s^{3}\over {\tilde g}}A \ ,\;\;\;\;\;\;\;\;\;\;\;\;\;\;\;\;
A_5={l_s\over {\tilde g}}A \ ,\cr
&&\cr
A_{015}&=&-\frac{l_s^{3}}{{\tilde g}}A \ ,
\;\;\;\;\;\;\;\;\;\;\;\;\;
A_{01234}=-\frac{l_s^{5}}{{\tilde g}}A \ ,\cr
&&\cr
B_{01}&=& l_s^2a_{\rm eff}^2u^2 \ ,\;\;\;\;\;\;\;\;\;\;\;
e^{2\phi}={\tilde g}^2
a^2_{\rm eff}u^2A^{-2} \ ,
\eea
where $A$ is given by (\ref{A}).

Apart from the the RR 3-form potentials, this system is the OD4 
theory studied in  
\cite{GMSS} and from the supergravity point of view in \cite{AOR}.
NS5-branes with one type of RR field on their worldvolume have been also considered in
\cite{PAP} and from the supergravity point of view in  
\cite{{ILPT},{CP},{OHZ}}. 

In the IR and UV regimes the dilaton is large and the 
good description is given by 
an eleven-dimensional supergravity background.
Consider the UV regime. The 
eleven-dimensional supergravity background
\bea
l_p^{-2}ds^2&=&{\hat h}^{-1/2}\left(\frac{u^{2/3}}{N^{1/3}(b_1^3b_2)^{1/2}}
(-dx_0^2+dx_1^2+{\hat h}_1d{\tilde x}_{2,3,4}^2+{\hat h}_2d{\tilde x}_{5}^2
\right. \nonumber \\ && \left.
+{N^{2/3}\over b_2^2u^{4/3}}{\hat h}^{-1}_2
(dx_6+a_2^2u^2{\hat h}_2d{\tilde x}_5)^2
+{{N^{2/3}b_2}\over {R u^{4/3}}}(du^2+u^2d\Omega_3^2)\right)\ , \cr
&&\cr
C_{234}&=&{l_p^3\over b_1^3}a_1^2u^2{\hat h}_1 \ ,\;\;\;\;\;\;\;\;\;\;\;\;\;\;\;
\;\;C_{015}={l_p^3\over b_2^3}a_2^2u^2{\hat h}_2 \ ,\cr
&&\cr
C_{016}&=&{l_p^3\over (b_1b_2)^{3/2}}a_1a_2u^2 \ ,\;\;\;\;\;\;\;\;
C_{012346}={l_p^6\over b_1^6}a_1^2u^2{\hat h}_1 \ ,
\label{11}
\eea
where ${\hat h}_i^{-1}=1+a_i^2u^2$. The rescaled coordinates are defined as
\be
{\tilde x}_{2,3,4}={b_1^{3/2}\over l_p^{3/2}}x_{2,3,4} \ ,\;\;\;\;\;\;\;\;
{\tilde x}_{5}={b_2^{3/2}\over l_p^{3/2}}x_{5} \ .
\label{11res}
\ee

This background can be considered from another point of view.
One can start with M5-branes with worldvolume coordinates $x_0,...,x_5$ and
smear in the direction $x_6$. One then rotates in $x_5,x_6$ plane.
With the addition of the $C$ fields one has 
\bea
ds^2&=&h_1^{-1/3}\left(f^{-1/3}(-dx_0^2+dx_1^2+h_1dx_{2,3,4}^2+h_2dx_5^2)+
f^{2/3}h_2^{-1}(dx_6+Adx_5)^2 \right. \nonumber \\ && \left.
+f^{2/3}(dr^2+r^2d\Omega_3^2)\right) \ ,\cr
&&\cr
C_{234}&=&\tan\theta_1f^{-1}h_1 \ ,\;\;\;\;\;\;\;\;\;\;\;
C_{015}=-\sin\theta_1\cos\theta_2f^{-1}h_2 \ ,\cr
&&\cr
C_{016}&=&\sin\theta_1\sin\theta_2f^{-1} \ , \;\;\;\;\;\;\;C_{012346}=-\sin\theta_2
\cos\theta_1f^{-1}h_1 \ ,\cr
&&\cr
A&=&\tan\theta_2f^{-1}h_2 \ ,\;\;\;\;\;\;\;\;\;\;\;f=1+{Nl_p^3\over \cos\theta_1
\cos\theta_2 R r^2} \ ,
\label{OMAN}
\eea
where $R$ is the radius of $x_6$, and 
$h^{-1}_i=\sin^2\theta_if^{-1}+\cos^2\theta_i$.
The decoupling limit is defined such that 
$l_p\rightarrow 0$  with 
\be
u={R^{1/2}r\over l_p^{3/2}} \ ,\;\;\;\;\;\;\;\;\;\;b^{3/2}_i={l_p^{3/2}\over 
\cos\theta_i} \ ,
\ee
and $R$ fixed.
This leads to the background (\ref{11}), (\ref{11res}).
Thus, we can think of (\ref{11}) OM theory \cite{GMSS} ``at an angle''.\\

\underline{{\bf $p=3$}}\\

When $p=3$, the supergravity background  is S-dual to the background 
describing  the NCSYN theory in (5+1)-dimensions
with four non-commutative directions. The phase structure and some properties
of this theory
have been studied in \cite{AOJ}.

\vskip 2cm  

{\bf Acknowledgement}:
We would like  to thank A. Giveon and B. Kol for valuable discussions.

\end{document}